# Direct epitaxial integration of the ferromagnetic semiconductor EuO with silicon for spintronic applications

Dmitry V. Averyanov, Peter E. Teterin, Yuri G. Sadofyev, Andrey M. Tokmachev,
Alexey E. Primenko, Igor A. Likhachev & Vyacheslav G. Storchak*

National Research Centre "Kurchatov Institute"
Kurchatov Sq. 1, Moscow 123182, Russia
*e-mail: mussr@triumf.ca

**Materials in which charge and spin degrees of freedom interact strongly offer applications known as spintronics. Following a remarkable success of metallic spintronics based on the giant-magnetoresistive effect, tremendous efforts have been invested into the less developed semiconductor spintronics, in particular, with the aim to produce three-terminal spintronic devices, *e.g.* spin transistors. One of the most important prerequisites for such a technology is an effective injection of spin-polarized carriers from a ferromagnetic semiconductor into a nonmagnetic semiconductor, preferably one of those currently used for industrial applications such as Si – a workhorse of modern electronics. Ferromagnetic semiconductor EuO is long believed to be the best candidate for integration of magnetic semiconductor with Si. Although EuO proved to offer optimal conditions for effective spin injection into silicon and in spite of considerable efforts, the direct epitaxial stabilization of stoichiometric EuO thin films on Si without any buffer layer has not been demonstrated to date. Here we report a new technique for control of EuO/Si interface on submonolayer level which may have general implications for the growth of functional oxides on Si. Using this technique we solve a long-standing problem of direct epitaxial growth on silicon of thin EuO films which exhibit structural and magnetic properties of EuO bulk material. This result opens up new possibilities in developing all-semiconductor spintronic devices.**

Modern information technology is based on the fundamental dichotomy: it utilizes charge of electrons to process information in semiconductors and their spin to store information in magnetic materials. Strong correlation of spin and charge degrees of freedom in the same material makes it possible to manipulate magnetically stored information with electric fields and/or modify fast logic gates by changing the magnetisation of their components. In metal multilayers, such effects are manifest in giant magnetoresistance, where the orientations of the macroscopic magnetisation in adjacent layers determine the electrical resistance of the structure [1,2]. Metallic spintronic devices, such as hard disk read heads and magnetic random access memory are among the most successful technologies of the past decades. However, metals cannot enhance signals – the prerequisite for transistor technology readily offered by semiconductors.

The development of semiconductor spintronics requires the ability to inject, modulate and detect spin-polarized carriers in a single device, preferably made of technologically important materials currently used in integrated circuits such as Si or GaAs [3,4]. Thus far, the spin of the carriers has played a minor role in semiconductor devices mainly because Si and GaAs are nonmagnetic. On the other hand, the enhanced spin-related phenomena realized in diluted magnetic semiconductors (DMS) (especially

GaMnAs films [5]) open the way for applications in spintronics [6]. The interplay between electrical and magnetic properties of III-V DMS has been reported in literature as electric field control of FM [7] and magnetic field driven giant Hall resistance jumps [8]. The successful demonstration of the injection of polarized spins in spintronic devices [9] involves these Mn-doped DMS as spin injectors. However, such doping significantly affects their homogeneity [10,11] which may cause strong spin-flip scattering of spin-polarized carriers making spin injection rather ineffective.

Recent years witnessed many attempts to inject spin-polarized electrons into Si [4] but large currents with high spin polarisation are yet to be demonstrated. Optical orientation of spins is inefficient in Si owing its indirect band gap. Perhaps the most promising strategy in creating spin-polarized carriers in conventional electronic devices is to integrate materials with high spin polarisation with Si *by means of electrical contact* (known as electrical spin injection). Here one needs to find a way for a) effective spin injection and b) effective maintenance of spin polarisation in the host material. Silicon has long been predicted a superior spintronic material for effective maintenance of spin polarisation having exceptionally long spin coherence lifetime ($\sim 10^{-5}$-$10^{-4}$ s) and spin-decoherence transport length ($\sim 1$ μm) due to very weak spin-orbit interaction and lattice inversion symmetry [12,13]. In contrast, much work on metallic FM spin injectors proved them to be ineffective in both ohmic regime and tunneling injection [3,4,14]. Perhaps the best choice for the FM injector in such devices is magnetic semiconductors (MS) because of their compatibility with nonmagnetic semiconductors (formation of a heterojunction): the use of MS as a spin-polarized carrier injector avoids the so-called "conductivity mismatch" [15] which presents a fundamental difficulty for effective spin injection into a semiconductor from the FM metals. The interfacial quality of the heterojunction may have a strong effect on the injection efficiency, so epitaxial growth will be necessary. Materials such as DMS Mn-doped Si [16], Mn-doped chalcopyrites [17] or MS EuO [18] have all been suggested but none as yet have been demonstrated as a spin injector for Si [19].

Being representatives of strongly correlated electron systems, MS demonstrate strong dependence of electrical and optical properties on the magnetisation and spin fluctuations of their magnetic lattice [20,21]. Concentrated MS offer several important advantages over DMS such as higher magnetisation, spatial magnetic homogeneity and wider range of conductivity tuning by doping, so that they can be used as spin filters in the insulating state and as spin injectors when doped [22,23]. Being doped, though, at high temperature, these materials typically enter into dominant states that are not spatially homogeneous due to formation of magnetic polarons – few-body systems comprised of electron and local magnetic moments of the host [20,24-27]. However, this unwanted formation does not take place when magnetisation of the lattice is significant, leaving the host material perfectly homogeneous in the region of its employment as a spin injector [25-27].

Owing its outstanding magnetic and transport properties among other MS, EuO has recently attracted much attention as having tremendous potential for semiconductor spintronics, in particular, when integrated with Si [18,28-30]. Not only does doped EuO exhibit a spin polarization close to 100% due to enormous (~0.6 eV) spin splitting of its conduction band but also it can be conductance-matched with Si by doping with oxygen vacancies or trivalent rare-earth atoms such as Gd, La or Lu [18,30-33]. Its Heisenberg-like magnetism (7 μ$_B$ per Eu$^{2+}$ ion) arises from the half-filled 4*f* states which constitute the top of the valence band [33]. Stoichiometric EuO has a low FM Curie temperature $T_c$ of about 69 K, but chemical doping and axial strain can increase $T_c$ significantly [32-36]. Its band gap of 1.12 eV matches that of Si [33]. In addition to their structural compatibility, EuO is the only binary magnetic oxide that exhibits thermodynamic stability in contact with Si [37]. Its remarkable bulk properties – metal-insulator transition accompanied by 13-15 orders of magnitude change in resistivity, colossal magnetoresistivity effect of about 6 orders of magnitude in a modest magnetic field of 2 T, exceptionally strong magneto-optics effects, a high sensitivity of transport and magnetic properties to doping [18,33] – plants firm expectation that the functionality of spin-selective ohmic EuO/Si devices [18] can be tuned as never before.

Despite the tremendous potential for spintronics, the epitaxial growth of single crystalline stoichiometric EuO films *directly* on Si has not been reported to date [28]. Thin films of EuO can be epitaxially grown on inert substrates like $YAlO_3$ [38], yttria-stabilized cubic zirconia (YSZ) [39] or MgO [40], especially when the lattice mismatch is small, but stabilisation of EuO on reactive surfaces of Si and GaAs [41] is notoriously difficult leading to polycrystalline EuO [28,38] and/or weak or no FM behaviour [41]. The breakthrough for integration of EuO with Si has been achieved by using an intermediate SrO buffer layer [18,42]. An intermediate layer, however, reduces the probability of spin-polarized carriers injection exponentially. In particular, an enormous band gap of about 6 eV in SrO makes injection of spin-polarized carriers into Si rather ineffective.

The continuing attempts to grow EuO/Si heterojunctions [28-30,43,44] are checked by the presence of large amounts of impurity phases at the interface. These phases are not only detrimental to the growth of EuO films; they also prevent spin injection due to spin-flip scattering. Thus, in order to create epitaxial structures in which the properties of the underlying silicon and the overlying EuO film both attain their full potential, there is a clear need for direct epitaxial integration making an ohmic contact between EuO and Si.

Here we report specific methods that make possible a direct epitaxial integration of stoichiometric single crystalline EuO films with Si. The structural and magnetic properties of our films rival those of bulk single crystals. We show that the control of the silicon/oxide interface is critical for the direct epitaxial growth and specific preparation of Si surface allows implementation of such control. In addition, we solve a long-standing problem of a capping layer for EuO films by controlled formation of a very thin (~2-3 nm) layer of $Eu_2O_3$ on the surface, which prevents further oxidation and/or hydration pretty much the same way as $Al_2O_3$ layer protects Al metal.

Interfacing EuO with silicon is a major challenge: in addition to the significant thermal and enormous (+5.6%) lattice mismatch it adds the complexity of joining covalent systems to ionic ones, chemical interactions at interfaces and possible compositional and structural changes. That is why we had a long way to go to grow epitaxial films of EuO directly on Si. Our first attempts followed a standard route for EuO growth. Adsorption controlled (distillation) regime is suitable for formation of high-quality EuO films on oxide substrates ($YAlO_3$, YSZ, MgO, *etc.*). This growth mode is based on larger than stoichiometric Eu flux to prevent formation of higher oxides, with excessive Eu atoms being evaporated. The latter process requires high substrate temperatures (above 400 °C). Exposure of Si substrate to high temperature, though, leads to formation of amorphous $SiO_2$ and metal bulk silicide $EuSi_2$ in oxidizing and reducing environment, respectively. These reactions are suppressed at lower temperatures but any departure from stoichiometry boosts formation of higher oxides ($Eu_3O_4$ and/or $Eu_2O_3$) or large number of vacancies ($EuO_{1-x}$). As a result, the range of the temperatures and ratios of Eu and $O_2$ fluxes is extremely narrow making selection of optimal growth parameters notoriously difficult. Obviously, it calls for protection of the Si surface.

The seminal paper by McKee *et al.* [45] has provided a robust solution to the long-standing problem of growing a commensurate single crystalline oxide interface with Si by employing intermediate sub-monolayer alkaline-earth silicide to protect the silicon surface from oxidation. The clean unreconstructed (100) silicon surface is formed by Si atoms with two singly occupied dangling bonds per atom. Pairs of silicon atoms form covalent chemical bonds leading to the dimer row (2×1) reconstruction, which consists of alternating (1×2) and (2×1) domains separated by single-atom height steps. It leaves one singly occupied dangling bond per atom making silicon surface highly reactive. Formation of surface alkaline-earth silicide is aimed at the saturation of these dangling bonds to exploit reactivity of the Si surface in a controlled fashion. Most of the studies involve strontium which is highly relevant to the growth of EuO because almost identical ionic radii of $Eu^{2+}$ and $Sr^{2+}$ lead to the well-known isomorphism of Eu(II) and Sr compounds with very close structural parameters.

The standard procedure for Sr-passivation of the Si surface includes creation of Sr adlayer with the coverage of 1/2 monolayer (ML) corresponding to the stoichiometry $SrSi_2$. Each metal atom donates 2 electrons and, therefore, each dangling bond becomes occupied by an electron pair. As a result, this structure has no surface states in the band gap of the host [46], which is rather important for applications. The presence of a band gap adds to the resistance of the surface to oxidation [47] although one should note that both theory [48] and experiments [49] predict the silicide to become partly oxidized with formation of M-O-Si bonds. This (2×1) reconstruction is routinely used for the growth of AO and $ABO_3$ compounds on Si, including EuO [18,30,42]. Both our results and other reports show the presence of impurity phases in such EuO/Si structures. It means that the protection of the surface is not sufficient for producing clean EuO/Si interfaces and other options should be explored.

It is known that adsorption of Sr (and similarly Eu) on the (100) silicon surface results in a phase diagram which involves a series of surface reconstructions depending on the temperature and the metal coverage, and they can be used in an attempt to improve the surface protection. The (2×3) structure corresponding to 1/6 ML metal coverage has a large amount of unsaturated dangling bonds and it is almost as reactive as pure Si. Indeed, an attempt to use this structure as a template for BaO leads to an amorphous growth [49]. Low- and high-temperature procedures to obtain the (2×1) reconstruction of the surface are very similar as long as the quality of the grown oxide is concerned [49].

Better results are expected from surface silicides with higher than 1/2 ML metal coverage of the surface. Our experiments show that Eu forms a stable surface silicide with the (5×1) reconstruction. Similar structures are obtained for other divalent metals Yb, Ca, and Sr [50]. It is known to have more than 1/2 ML metal coverage but the actual value is debated in the literature. Higher metal coverage should provide better protection of the surface. What is most important, it does not lead to surface states in the band gap of silicon because the structure is expected to be formed by breaking Si-Si dimers and saturating the resulting free valences with metal atoms. This can be confirmed by electron spin resonance (ESR) measurements which proved to be extremely sensitive to the presence of unpaired electrons (dangling chemical bonds). Indeed, we do not observe any ESR signal for Si protected by surface europium silicide for both structures with (2×1) and (5×1) reconstructions.

The conclusion about better protection of the surface by silicides with higher metal content has some experimental backing: the authors of Ref. [48] mention that among four different reconstructions of Ba and Sr on silicon used to grow lattice-matched $Ba_{0.7}Sr_{0.3}O$ the best results are obtained for the (5×1) reconstruction of Sr on Si. Although this result cannot be directly transferred to the growth of EuO on Si due to the large lattice mismatch and high reactivity of EuO, it is a strong indication that the use of the (5×1) reconstruction may add crucial advantages to the growth process. Our results below show that this improved protection of the Si surface leads to direct epitaxial integration of EuO with silicon.

To form a protective submonolayer Eu silicide the reconstructed bare Si surface is exposed to Eu flux. The increasing Eu coverage leads to a series of surface phases as witnessed by reflection high-energy electron diffraction (RHEED): (2×1) + (1×2) Si reconstruction is transformed into (2×3) + (3×2), then into (1×2) + (2×1) and finally into (1×5) + (5×1) (Fig. 1). Periodic symmetries (m×n) and (n×m) are both present for surfaces with steps of single-atom height. Similar to Sr surface silicides, for which experiments with Si wafers having doubled height of the steps have been carried out [49], the sequence of transformations is (2×1) Si into (2×3), (1×2) and then into (1×5) silicide phase. The resulting state of the surface depends on the fine balance of Eu adsorption and desorption, which is tuned by substrate temperature and Eu flux. The (1×5) + (5×1) surface silicide phase is formed at 660 °C and remains stable at lower temperatures.

To grow thin EuO films on thus protected Si surfaces we developed a two-step protocol. The first step is a low-temperature growth (stage A) in close to stoichiometric regime with slight excess of Eu atoms to avoid formation of $SiO_2$ and bulk $EuSi_2$. It requires fine tuning of the substrate temperature and fluxes. We find that 10 MLs of EuO are enough to prevent formation of unwanted phases. Changes of the EuO lateral lattice parameter during EuO growth can be determined from the evolution of the distance between streaks in corresponding RHEED patterns. The very beginning of epitaxial growth follows the Stranski-Krastanov regime with wetting layer thickness close to 1 ML. Fig. 2 shows that highly stretched layers of EuO become progressively more relaxed and the lattice parameter changes from the value corresponding to Si (5.44 Å) toward that of bulk EuO (5.14 Å). The changes are not uniform: the first 4 MLs exhibit the largest lattice relaxation; 10 MLs of EuO on Si guarantee the full lattice relaxation. Because of considerable lattice mismatch of EuO and Si, one expects growth of 3D islands, followed by pseudomorphic growth, which are typical for the Stranski-Krastanov regime. Instead, we do not see any form of 3D growth – the intensity of RHEED reflections is modulated rather slightly along the streaks, but, what is most important, typical spots characteristic to 3D growth are not detected.

The grown film is stable up to 510 °C at least, and it can be annealed at that temperature to improve the crystalline quality. The second step (stage B) is the growth at higher temperature (470 °C), where it is controlled by Eu distillation. This step is similar to the growth of EuO on oxide substrates. It is followed by annealing at 530 °C. Typical RHEED image of EuO film grown this way is shown in Fig. 3 and consists of streaks, indicating that the surface is smooth. Observation of clean and sharp Kikuchi lines also points at the crystalline surface. It is also significant that proposed recipe yields stable reproducible results: grown films have identical properties. One can notice that the two-step procedure reminds growth of EuO on Si with protective buffer layer of SrO. The important difference is that in our case EuO is grown in direct contact with Si making efficient spin injection possible.

EuO is highly reactive and when exposed to the atmosphere it forms hydroxide, higher oxides and even carbonate. Degradation of unprotected EuO thin films is very fast. Therefore, EuO thin films need protection. Different materials have been proposed and used as capping layers: Si, Ti, Al, $SiO_x$, $Al_2O_3$, *etc.* Our experiments show that thick enough capping layers formed by (oxidized) Al and $SiO_x$ provide sufficient protection. We propose a more advanced way to protect EuO films without introducing new components to the system. It is based on *controlled* oxidation of the surface of EuO with formation of inert $Eu_2O_3$. EuO films were exposed to a low oxygen flux in the vacuum chamber. The initially bright RHEED pattern extinguishes so that EuO reflections, although strongly obscured, are still visible. EuO films protected by $Eu_2O_3$ layer show no signs of degradation after months-long exposure to the air: both x-ray diffraction pattern and visual appearance remain unchanged. X-ray reflectivity measurements estimate the thickness of the protective layer to be about 2-3 nm.

X-ray diffraction measurements of EuO films are shown in Fig. 4. In contrast to other works, our θ-2θ diffraction scan (Fig. 4(a)) demonstrates peaks from (002), (004) and (006) reflections without any traces of undesired phases. All the peaks correspond to the same orientation of EuO and Si atomic planes. Maximal intensities of EuO (00n) and Si (00n) XRD reflections are observed for the same orientation of the sample, demonstrating that the angle between lateral atomic planes is negligible. Location of EuO reflections in the θ-2θ diffraction pattern corresponds to the lattice parameter *a* =5.1393±0.0001 Å, which is somewhat smaller than the bulk value (5.1435 Å). One should notice that this lattice parameter corresponds to the direction orthogonal to EuO/Si interface. The difference is probably caused by incomplete relaxation of the film: it is laterally stretched and, hence, some vertical compression is expected to avoid significant changes of the unit cell volume. Full coincidence of EuO and Si peaks on (202) reflection φ-scan (Fig. 4(b)) indicates that vertical facets of fcc structures of both systems are also aligned in parallel. Hence, the grown EuO films are epitaxial. Well-developed thickness fringes are observed for EuO (002), (004) and (006) reflections. Inset of Fig. 4(a) shows them for EuO (002) reflection. This characteristic feature of the x-ray diffraction is a result of the wave interference due to reflections at the interfaces, both top and bottom. Taking into account the small value of the x-ray wave length (1.5418 Å), the observation of the thickness fringes is a fingerprint of atomically sharp interfaces; otherwise the reflected waves cannot maintain the coherence.

The stoichiometry of the grown EuO films was controlled by Rutherford backscattering (RBS). A characteristic RBS spectrum of the EuO films is shown in Fig. S1 (see Supplementary Information). The best fit corresponds to stoichiometric EuO. The calculated thickness of the films corresponds perfectly to the values determined from the periods of XRD thickness fringes. RBS spectra exhibit strong channeling – yet another indication of the epitaxial integration of EuO and Si. It is evident from the spectra that the amplitude of the Eu peak diminishes when the incident ion beam is aligned along the [001] axis of the Si substrate.

Magnetic measurements (using superconducting quantum interference device) of both the fully grown EuO films (stages A and B) and the first low-temperature 10 MLs (stage A) exhibit the onset of ferromagnetism. Fig. 5(a) shows temperature dependence of the in-plane DC magnetisation of the fully grown EuO film in external magnetic field H=1 Oe. Temperature dependence of the AC (1 Hz frequency) magnetic susceptibility of the same EuO film in external magnetic field H=0 is presented in Figure 6. The Curie temperature is determined from both DC and AC measurements to amount 68±1 K – the same as in the bulk EuO. Any significant amount of defects would shift $T_c$ above the error bars. (Our close to stoichiometry EuO films are insulating, which prevents conventional measurements of their transport properties due to shunting by Si substrate.) The field dependence of in-plane DC magnetisation measured at 2 K (inset of Fig. 5(b)), shows hysteresis with coercivity ~ 90 Oe. The saturation magnetisation (Fig. 5(b)) per Eu atom is determined to be 6.9±0.1 $\mu_B$. This is consistent with the best bulk samples and corresponds to $4f^7$ configuration of $Eu^{2+}$.

10 ML-thick films show somewhat decreased temperature of the ferromagnetic transition (see Fig. S2 in Supplementary Information). This is not that surprising as spontaneous magnetisation of the Heisenberg ferromagnet EuO is caused by interatomic exchange interaction which strongly depends on the distance between Eu ions. Our RHEED measurements of the lattice parameter changes during stage A of the growth clearly show that EuO is laterally stretched (Fig. 2). As a consequence, Eu-Eu distances become larger and, hence, the Curie temperature becomes smaller than in the bulk.

In summary, we provide a recipe for epitaxial integration of EuO directly with silicon. We believe that there are two key ingredients, which made the solution possible. First, protection of the Si surface with the surface silicide having a larger metal content; in our case it corresponds to (5×1) reconstruction. Clearly, this recipe is not limited to europium silicide and the growth of EuO, it can be used for protection of the Si surface by strontium silicide and improved growth of other oxide materials, including complex ones. Second, the developed two-step growth protocol with the low-temperature stoichiometric regime preceding the adsorption-controlled growth ensures sharp and uniform interfaces without any impurity phases being detected. The proposed procedure of EuO protection by a layer of $Eu_2O_3$ seems to be clean and easy to implement. The grown thin films of EuO thus become an ideal candidate for successful injection of spin-polarized electrons into Si. We are currently exploring this possibility – the results will be published elsewhere. We hope that our studies open up a viable route to the ultimate goals of semiconductor spintronics.

**Materials and Methods**

Thin films of EuO have been grown by molecular-beam epitaxy (MBE) in Riber Compact 12 system modified for the growth of oxides. The background pressure is less than $10^{-10}$ Torr. Eu (99.99 % purity), $O_2$ (99.9995 % purity), $SiO_x$ (99.99 % purity) and Al (99.999 % purity) were used as precursors. All materials were evaporated from BN effusion cells. High-ohmic Si (001) wafers with 0.5° miscut angle were used as substrates. Cell and substrate temperatures were measured by thermocouples. Approximate cell temperatures were: $T_{Eu}$~500 °C, $T_{SiOx}$=950 °C, $T_{Al}$=950°C. Substrate temperatures above 270 °C were also controlled by an optical infrared pyrometer with the working wavelength 0.9 μm. RHEED was used for *in situ* control of the crystalline and morphological state of the surface. The measurements were made with use of kSA 400 RHEED system, k-Space Associates, Inc. Fluxes of all the reactants were meticulously controlled by hot cathode beam flux monitor. Oxygen pressure was also monitored *in situ* by mass spectrometer.

To remove the natural amorphous $SiO_2$ layer substrates were heated up to 950 °C where $SiO_2$ is volatile. The resulting RHEED pattern clearly demonstrates formation of the Si (2×1) + (1×2) reconstruction – dimerization of surface Si atoms. After that stage, the substrate was cooled down to the temperature 660 °C and the Eu cell shutter was open. Formation of Eu on Si structure with (1×5) + (5×1) reconstruction takes approximately half a minute for the Eu flux about $5 \cdot 10^{-8}$ Torr. The low-temperature EuO growth started with simultaneous supply of Eu and oxygen. The ratio of Eu and oxygen fluxes requires fine tuning to maintain the film stoichiometry. It was performed step by step on the basis of approximate hot cathode beam flux monitor values, with further directions provided by RHEED-patterns. Thus grown 10 MLs of the EuO thin film were annealed for 15 min at 510 °C. The RHEED image at the end of this procedure is shown in Fig. S3. XRD-pattern of the same sample capped by $SiO_x$ for *ex situ* studies is also shown in Fig. S4. Results of magnetic measurements are presented in Fig. S2. High-temperature distillation stage of EuO growth was performed at 470 °C, the typical speed of the growth is kept at about 40 nm per hour. The following annealing at 530 °C was aimed at improvement of crystalline quality and decrease of the number of point defects. Capping was carried out at room temperature. Al or $SiO_x$ were evaporated from effusion cells while $Eu_2O_3$ protective layers were formed by exposing samples to an oxygen flux ($5 \cdot 10^{-8}$ Torr).

Magnetic measurements were performed with SQUID Quantum Design MPMS XL-7. X-ray diffraction spectra were recorded with use of Bruker K8 Advance and Rigaku SmartLab 9kW spectrometers with CuKα radiation. Rutherford backscattering spectra were measured for He ions with the energy 1.7 MeV.

**Acknowledgements**

The authors are grateful to S. Gudenko for ESR measurements. The work was financially supported by NRC "Kurchatov Institute", Russian Foundation for Basic Research (RFBR) through grant 13-07-00095, and Russian Science Foundation through grant 14-19-00662.

**Figure 1.** RHEED images along the [110] azimuth of reconstructed surface phases: (a) (2×1) + (1×2) Si; (b) (2×3) + (3×2) Eu on Si; (c) (1×2) + (2×1) Eu on Si; (d) (1×5) + (5×1) Eu on Si.

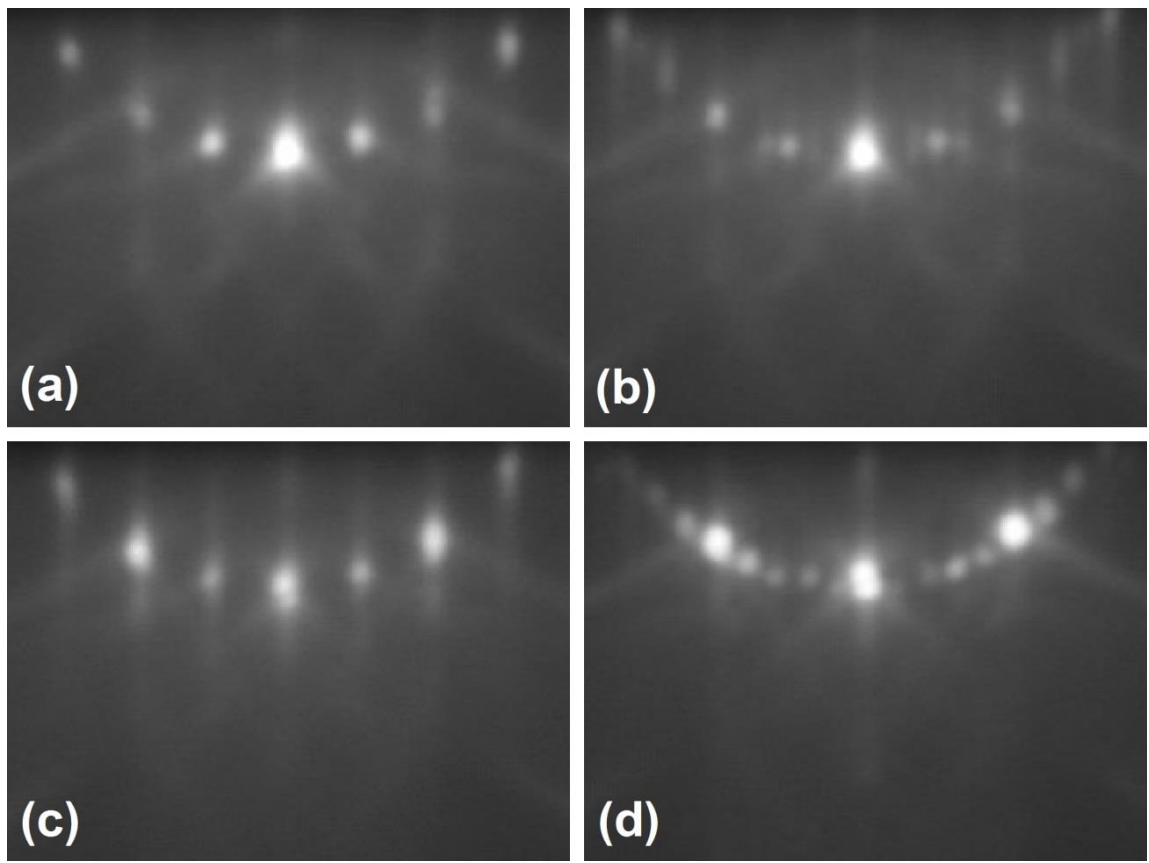

**Figure 2.** Changes of the lateral lattice parameter at the low-temperature (stage A) of EuO growth as determined by distance between reflections in the RHEED pattern. After growth of 10 MLs the film is completely relaxed. XRD measurements of thick EuO films also show that the vertical lattice parameter is very close to the bulk value.

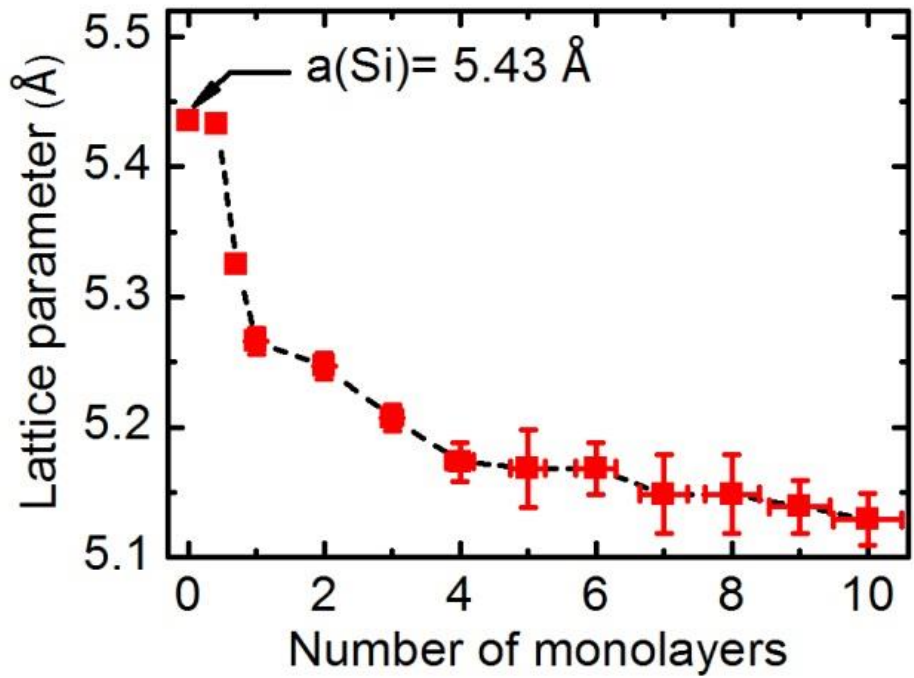

**Figure 3.** (a) Typical RHEED image along the [110] azimuth of the fully (according to the two-step protocol) grown $d_{EuO}$=40 nm thick EuO film, (b) corresponding horizontal intensity profile.

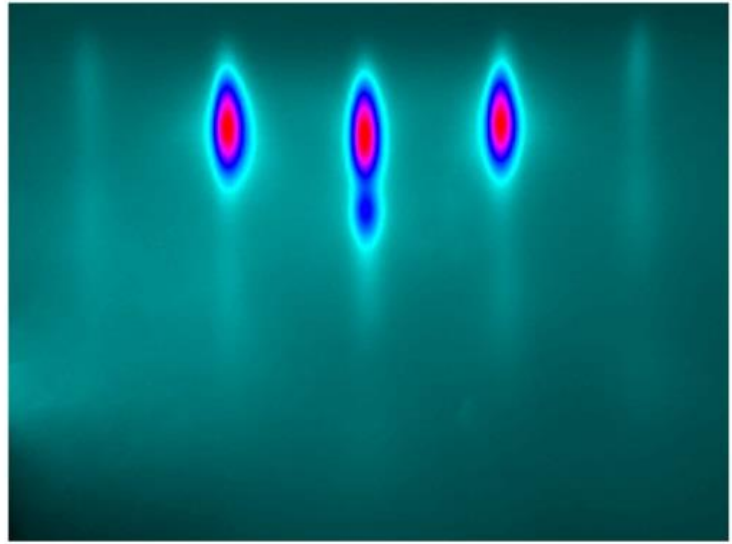

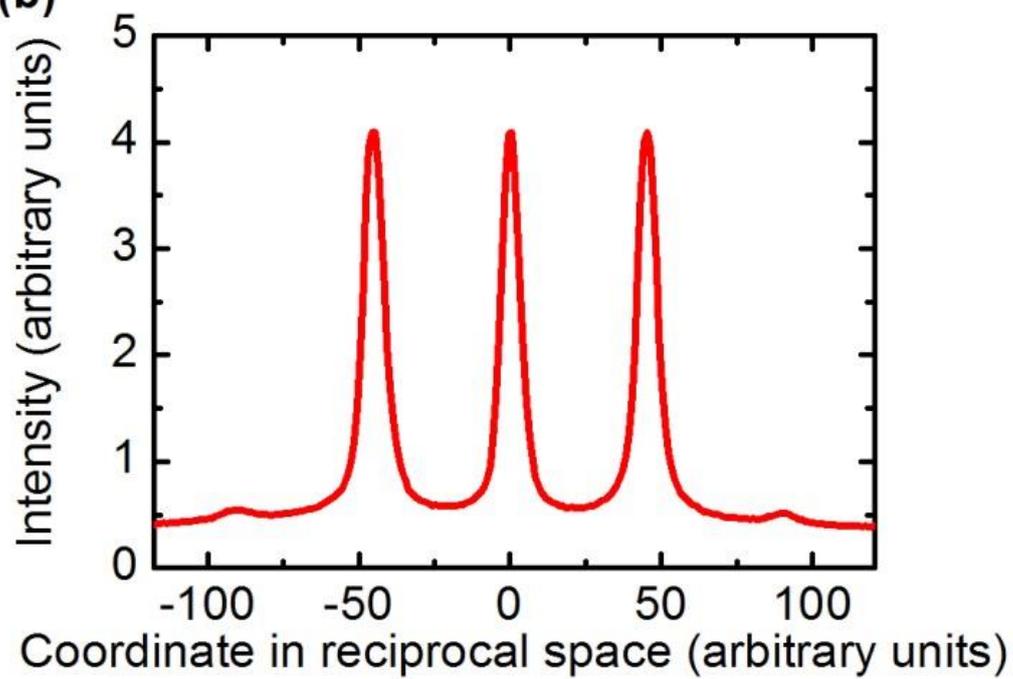

**Figure 4.** X-ray diffraction scans of SiO$_x$/EuO/Si sample: (a) θ-2θ diffraction spectrum; inset: thickness fringes of EuO (002) peak, (b) φ-scans of reflections for EuO (202) (blue) and Si (202) (red).

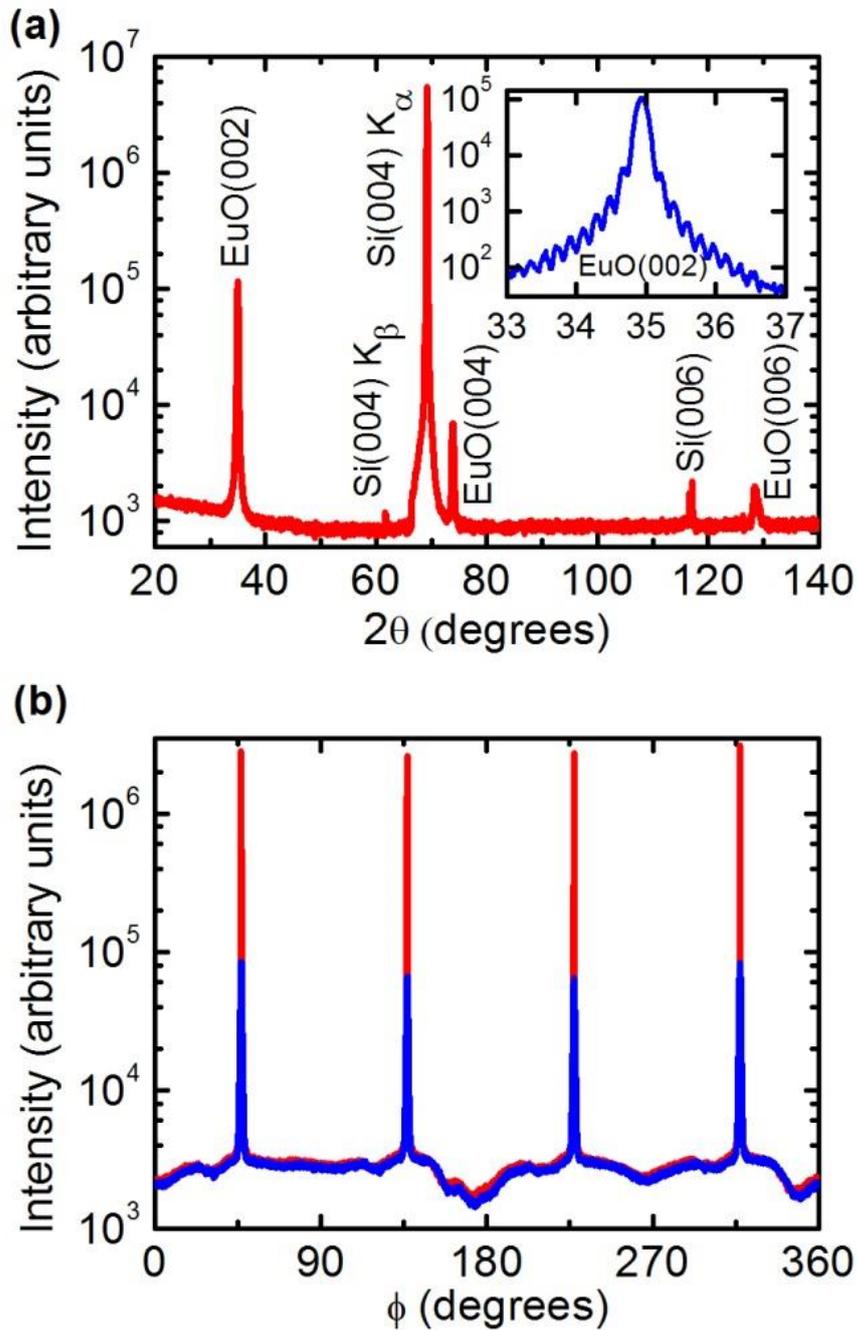

**Figure 5.** (a) Temperature dependences of the DC magnetisation of epitaxial $SiO_x/EuO/Si$ heterostructure ($d_{EuO}$=40 nm): warming zero-field-cooled curve (red) and cooling field-cooled curve (blue) in magnetic field H=1 Oe applied parallel to the film surface. (b) In-plane magnetization of epitaxial EuO film on Si as a function of the applied magnetic field H||[100] at T=2 K. Inset shows hysteresis loop.

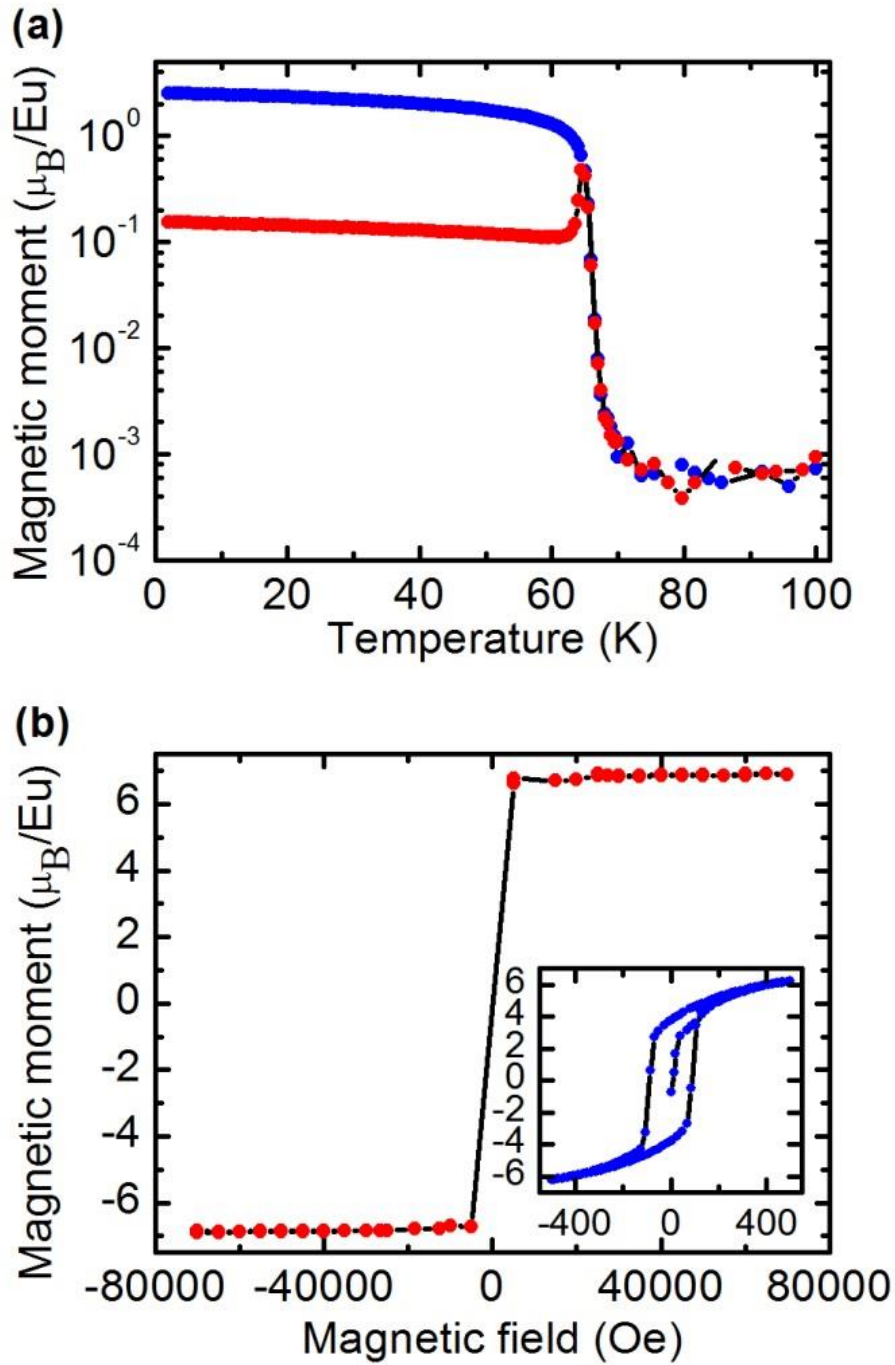

**Figure 6.** Temperature dependence of the AC (1 Hz) magnetic susceptibility of epitaxial SiO$_x$/EuO/Si structure (d$_{EuO}$=40 nm) in zero external magnetic field.

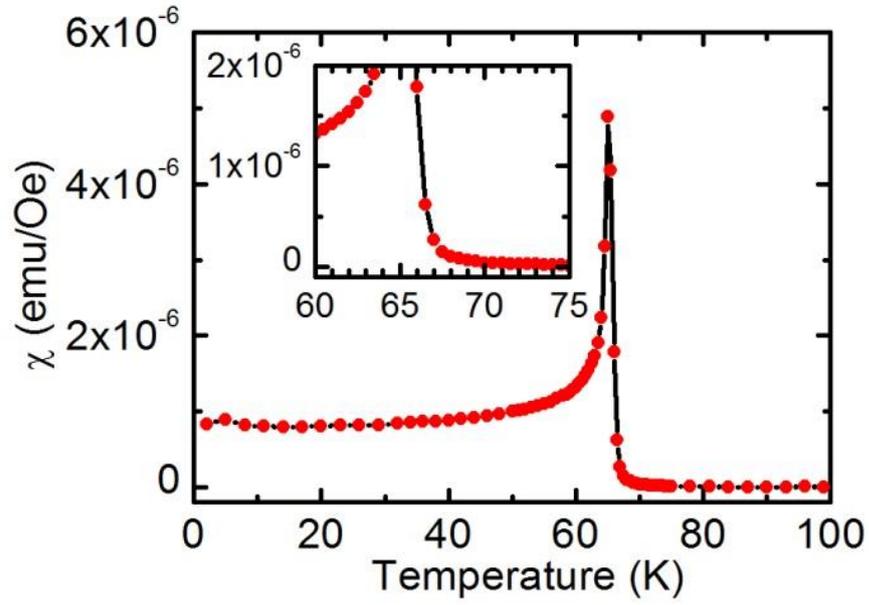

**Direct epitaxial integration of the ferromagnetic semiconductor EuO with silicon for spintronic applications**

**Supplementary information**

Dmitry V. Averyanov, Peter E. Teterin, Yuri G. Sadofyev, Andrey M. Tokmachev,
Alexey E. Primenko, Igor A. Likhachev & Vyacheslav G. Storchak

National Research Centre "Kurchatov Institute"
Kurchatov Sq. 1, Moscow 123182, Russia

**Figure S1.** Simulated RBS and channelling spectra of $Eu_2O_3$/EuO/Si heterostructure. Thickness of EuO layer $d_{EuO}$=70 nm. Protective $Eu_2O_3$ layer is not detected because of its low thickness.

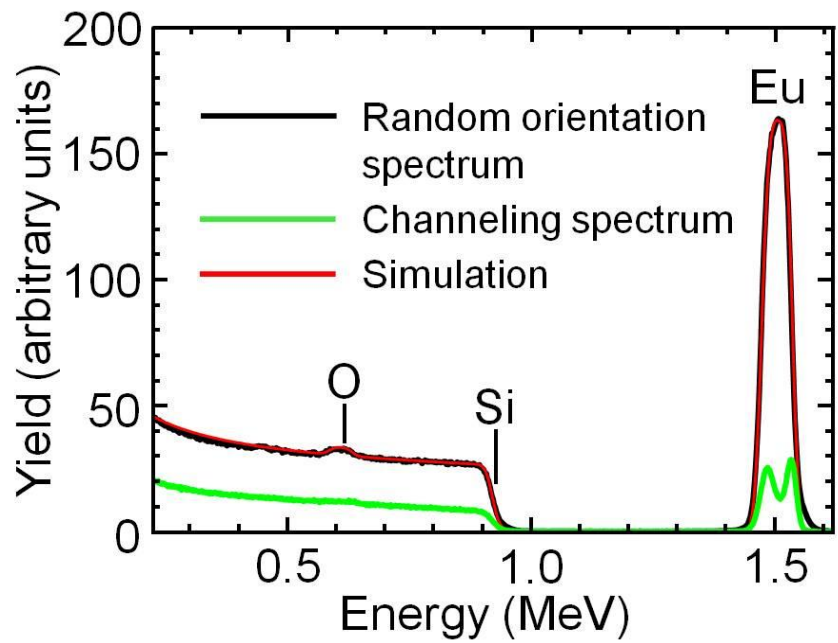

**Figure S2.** Temperature dependence of the magnetisation of 10 MLs EuO (stage A): a) warming zero-field-cool curve (red) and field-cool cooling curve (blue) in magnetic field H=10 Oe applied parallel to the film surface; (b) in-plane magnetisation as a function of the applied magnetic field H||[100] at T=2 K.

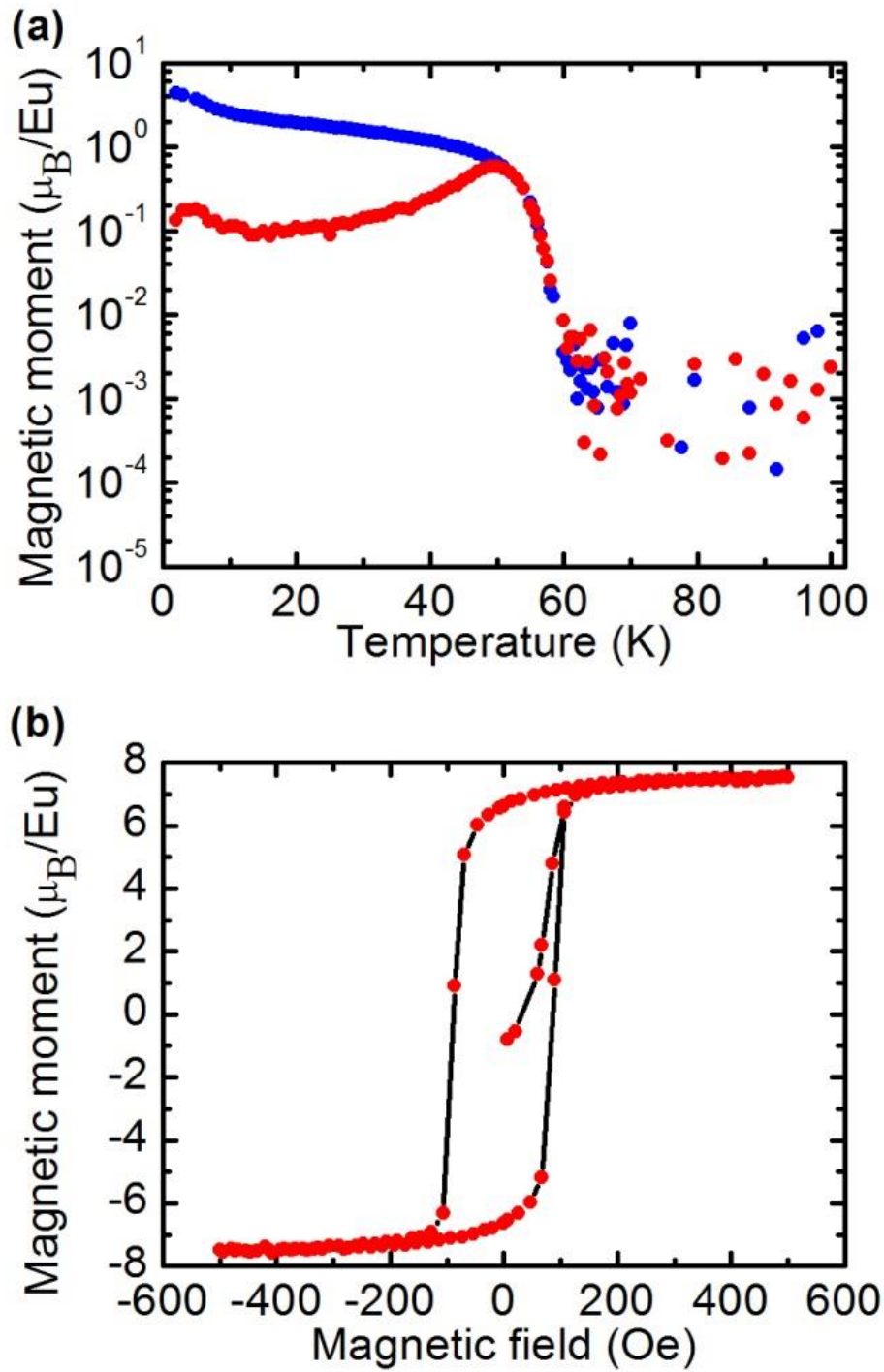

**Figure S3.** Typical RHEED image recorded along the [110] azimuth of the EuO film at the end of the low-temperature stage (stage A) of the growth.

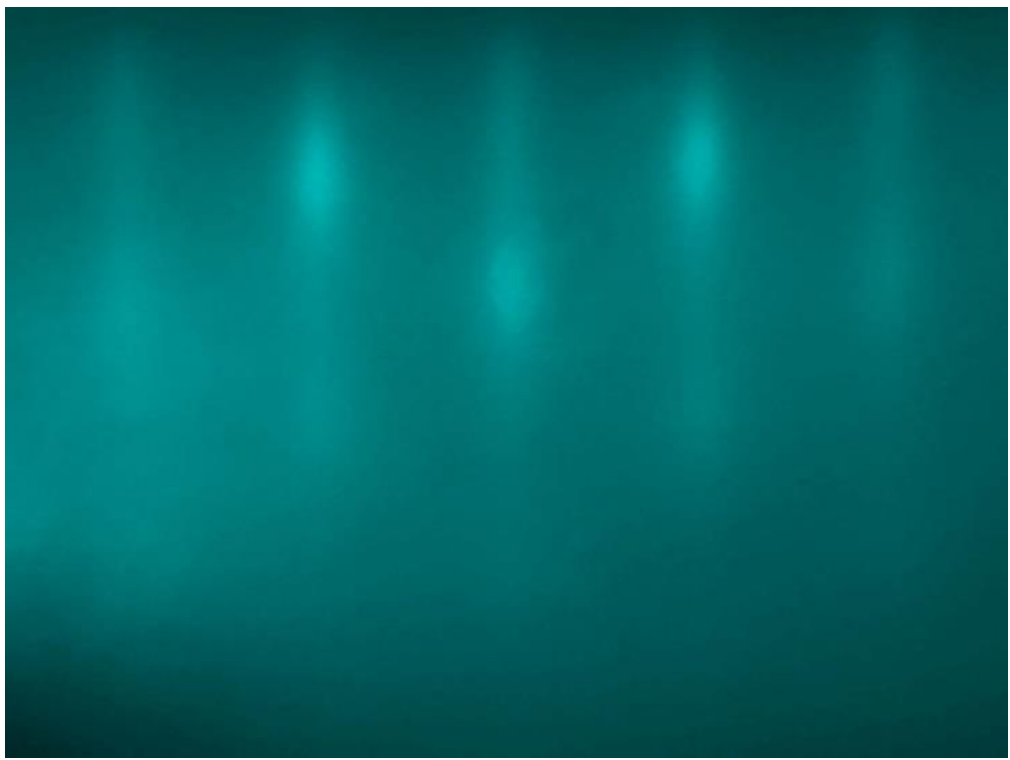

**Figure S4.** θ-2θ X-ray diffraction pattern of the sample $SiO_x$/EuO/Si. EuO thickness is 10 MLs (stage A).

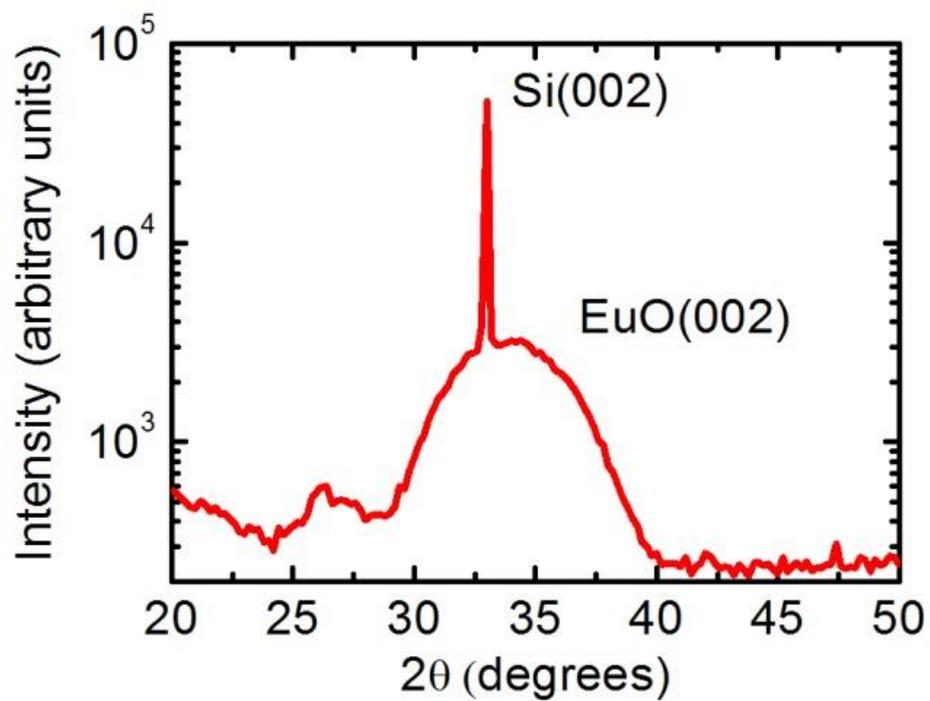